\documentstyle[12pt]{article}
\begin{document}
\title{GYROSCOPE PRECESSION IN CYLINDRICALLY SYMMETRIC SPACETIMES}
\author{L. Herrera$^1$\thanks{Also at Deapartamento de Fisica,
 Facultad de Ciencias,UCV, Caracas,
Venezuela; e-mail: lherrera@gugu.usal.es}, F. M. Paiva$^2$\thanks{e-mail:
fmpaiva@symbcomp.uerj.br} and N. O. Santos$^3$\thanks{e-mail:
nos@conex.com.br}
\\ \\
{\small $^1$Area de Fisica Teorica, Facultad de Ciencias,}\\
{\small Universidad de Salamanca, 37008 Salamanca, Spain.}\\
{\small $^2$Departamento de F\'{\i}sica Te\'orica, Universidade do Estado do
Rio de Janeiro}\\
{\small Rua S\~ao Francisco Xavier 524, 20550-013 Rio de Janeiro, RJ,
Brazil.}\\
{\small $^3$Laborat\'orio de Astrof\'{\i}sica e Radioastronomia}\\
{\small Centro Regional Sul de Pesquisas Espaciais - INPE/MCT}\\
{\small Cidade Universit\'aria, 97105-900 Santa Maria RS, Brazil.}}
\maketitle
\begin{abstract}
We present calculations of gyroscope precession in spacetimes described by
Levi-Civita and Lewis metrics, under different circumstances. By doing so we
are able to establish a link between the parameters of the metrics and
observable quantities, providing thereby a physical interpretation for those
parameters, without specifying the source of the field.
\end{abstract}
\newpage
\section{Introduction}
As stressed by Bonnor \cite{Bonnor} in his review on the physical
interpretation of vacuum solutions of Einstein's field equations, {\it
relativists have not been diligent in interpreting} such solutions. One way
to palliate this deficiency of the physical content of the theory consists
in providing  link between the characteristic parameters of the solutions
and
quantities measured from well defined and physically reasonable
experiments, providing thereby physical interpretation for those parameters.\par
It is the purpose of this work to establish such a link for cylindrically
symmetric spacetimes. The general form of the metric  in this case was
given by
Lewis \cite{Lewis,Kramer} and describes a stationary spacetime.This metric
can be split into
two families called Weyl class and Lewis class.Here we shall restrict our
study to the Weyl class where
 all parameters appearing in the metric are real. For the Lewis class these
parameters can be complex.
The corresponding static
limit was obtained by Levi-Civita \cite{Levi}.\par
The motivation for this choice is provided, on one hand, by the fact that
the physical interpretation
of the parametrers of these metrics
is still a matter of discussion (see \cite {Bonnor}), and on the other, by
the fact that some of the parameters of
these metrics are related to topological defects not entering  into the
expression of the physical components of curvature tensor.
 The physical({\it gedanken}) experiment proposed here consists in
observing the precession of a gyroscope under different conditions in such
spacetimes. Specifically, we calculate the rate of rotation of a gyroscope
at rest in the frame in which the metric is presented, and also, the total
precession per revolution of a gyroscope circumventing the symmetry axis,
along a circular path (geodesic or not). By doing so, the four parameters of
the Weyl class, from the Lewis metric, become {\it measurable}, in the sense
that they are expressed through quantities obtained from well defined and
physically reasonable experiments (we are of course not discussing about the
actual technical feasibility of such experiments).\par
All calculations are carried out using the method proposed by Rindler and
Perlick \cite{Rindler}, and a very brief resume of which is given in the
next section together with the notation and the specification of the
spacetime under consideration.In section 3 we calculate the rate of
precession of a gyroscope at rest in the original lattice, and in section 4
we obtain the precession per revolution relative to the original frame, of a
gyroscope rotating round the axis of symmetry. Finally the results are
discussed in the last section.
\section{The spacetime and the Rindler-Perlick \newline method}
\subsection{The Lewis metric}
The Lewis metric \cite{Lewis,Kramer} can be written as
\begin{equation}
ds^2=-fdt^2+2kdtd\phi+e^\mu(dr^2+dz^2)+ld\phi^2,
\end{equation}
where
\begin{eqnarray}
f=ar^{1-n}-\frac{c^2}{n^2a}r^{1+n},\\
k=-Af,\\
l=\frac{r^2}{f}-A^2f,\\
e^\mu=r^{(n^2-1)/2},
\end{eqnarray}
with
\begin{equation}
A=\frac{cr^{1+n}}{naf}+b.
\end{equation}
Observe that taking $ds$ in (1) to have dimension of length $L$ then
\begin{equation}
[t]=[L]^{2n/(1+n)}.
\end{equation}
\begin{equation}
[r]=[L]^{2/(1+n)}.
\end{equation}
and
\begin{equation}
[b]=[L]^{2n/(1+n)}.
\end{equation}
\begin{equation}
[c]=[L]^{-2n/(1+n)}
\end{equation}
whereas $n$ and $a$ are dimensionless and $e^\mu$ is multiplied by a unit
constant with dimensions
\begin{equation}
 [L]^{(2n-n^2-1)/(n+1)}.
\end{equation}

The four parameters $n,a,b$ and $c$ can be either real or complex, and the
corresponding solutions belong to the Weyl or Lewis classes respectively.
Here we restrict our study to the Weyl class (not to confound with Weyl
metrics representing
static and axially symmetric spacetimes).\par
The transformation \cite{Stachel}
\begin{eqnarray}
d\tau=\sqrt{a}(dt+bd\phi),\\
d\bar{\phi}=\frac{1}{n}[-cdt+(n-bc)d\phi],
\end{eqnarray}
casts the Weyl class of the Lewis metric into the Levi-Civita metric (for
recent discussions on these
metrics see \cite{Silva}-\cite{Herrera}, and references therein).
However the transformation above is not valid globally, and therefore both
metrics are equivalent only locally, a fact that can be verified by
calculating the corresponding Cartan scalars \cite{Silva}. In order to
globally transform the Weyl class of the Lewis metric into the static
Levi-Civita metric, we have to make $b=0$.
Indeed, if $b=0$ and $c$ is different from zero, (12) gives an admissible
transformation for the time coordinate and (13) represents a transformation
to a rotating frame.
However, since rotating frames  (as in special relativity) are not expected
to cover the whole  space-time and furthermore
since the new angle coordinate ranges from $- \infty$  to $\infty$, it has
been argued in the past \cite{Silva}
that  both
$b$ and $c$ ,have to vanish  for (12) and (13) to be globally valid. This
point of view is also  reinforced
by the fact that, assuming that
only $b$ has to vanish in order to globally cast (1) into Levi-Civita, we
are lead to the intriguing result that there is not dragging
outside rotating cylinders
(see comments at the end of section 3).We shall recall this question later.

\subsection{The Rindler-Perlick method}
This method consists in transforming the angular coordinate $\phi$ by
\begin{equation}
\phi=\phi^{\prime}+\omega t,
\end{equation}
where $\omega$ is a constant. Then the original frame is replaced by a
rotating frame. The transformed metric is written in a canonical form,
\begin{equation}
ds^2=-e^{2\Psi}(dt-\omega_idx^i)^2+h_{ij}dx^idx^j,
\end{equation}
with latin indexes running from 1 to 3 and $\Psi, \omega_i$ and $h_{ij}$
depend on the spatial coordinate $x^i$ only (we are omitting primes). Then,
it may be shown that the four acceleration $A_\mu$ and the rotation three
vector $\Omega^i$ of the congruence of world lines $x^i=$constant are given
by \cite{Rindler},
\begin{eqnarray}
A_\mu=(0,\Psi_{,i}),\\
\Omega^i=\frac{1}{2}e^\Psi(\det h_{mn})^{-1/2}\epsilon^{ijk}\omega_{k,j},
\end{eqnarray}
where the comma denotes partial derivative. It is clear from the above that
if $\Psi_{,i}=0$, then particles at rest in the rotating frame follow a
circular geodesic. On the other hand, since $\Omega^i$ describes the rate of
rotation with respect to the proper time at any point at rest in the
rotating frame, relative to the local compass of inertia, then $-\Omega^i$
describes the rotation of the compass of inertia (the {\it gyroscope}) with
respect to the rotating frame. Applying (14) to the original frame of (1),
with $t=t^{\prime},r=r^{\prime}$ and $z=z^{\prime}$, we cast (1) into the
canonical form (15), where
\begin{eqnarray}
e^{2\Psi}=f-\omega^2l-2\omega k,\\
\omega_i=(0,0,\omega_\phi),\\
\omega_{\phi}=e^{-2\Psi}(\omega l+k),\\
h_{rr}=h_{zz}=e^\mu,\\
h_{\phi\phi}=l+e^{2\Psi}\omega^2_{\phi}.
\end{eqnarray}
From (12)-(17) and
\begin{equation}
\Omega\equiv(h_{ij}\Omega^i\Omega^j)^{1/2},
\end{equation}
we obtain that
\begin{equation}
\Omega=[h_{zz}(\Omega^z)^2]^{1/2},
\end{equation}
where
\begin{equation}
(\Omega^z)^2=\frac{e^{4\Psi}\omega^2_{\phi,r}}{4e^{2\mu}(fl+k^2)}.
\end{equation}
\subsection{Circular geodesics}
From (18) we obtain with the condition $\Psi_{,i}=0$ that
\begin{equation}
\omega=\frac{-k_{_r}\pm(f_{,r}l_{,r}+k^2_{,r})^{1/2}}{l_{,r}},
\end{equation}
which yields the expression for the angular velocity of a particle on a
circular geodesic in Lewis metric (see (49) in \cite{Herrera}). Now
substituting (2)-(5) into (26) we obtain \cite{Mashhoon}
\begin{equation}
\omega=\left(\pm\omega_0+\frac{c}{n}\right)\left[1-b\left(\pm\omega_0+\frac{
c}{n}\right)\right]^{-1},
\end{equation}
where $\omega_0$ is the angular velocity when the spacetime is static,
$b=c=0$, given by Levi-Civita's metric,
\begin{equation}
\omega_0^2=\frac{1-n}{1+n}a^2r^{-2n}.
\end{equation}
We can calculate the tangential velocity $W$ of the circular geodesic
particles (see (53) in \cite{Herrera}),
\begin{equation}
W=\frac{\omega(fl+k^2)^{1/2}}{f-\omega k}.
\end{equation}
Substituting (2)-(5) and (28) into (29), we obtain
\begin{equation}
W=\left(\pm\omega_0+\frac{c}{n}\right)\left(ar^{-n}\pm\frac{c}{na}\omega_0r^
n\right)^{-1}.
\end{equation}
It is worth noticing the fact, which follows from (27) and (30), that $b$
and $c$ affect
the angular velocity $\omega$, while for the tangential velocity $W$ only
$c$ plays a role.
\section{Precession of a gyroscope at rest in the original latice}
To calculate the precession in this case, we only have to put $\omega=0$ in
(17)-(24) (see \cite{Bonnor1} for a similar case). Then, we obtain after
simple calculations,
\begin{equation}
\Omega=\frac{e^{-\mu/2}}{2f}\frac{|kf_{,r}-fk_{,r}|}{(fl+k^2)^{1/2}},
\end{equation}
or, using (2)-(5) in (31),
\begin{equation}
\Omega=cr^{(1-n^2)/4}\left(ar^{1-n}-\frac{c^2}{n^2a}r^{1+n}\right)^{-1}.
\end{equation}
Thus the parameter $c$, appears to be essential in the precession of a
gyroscope at rest in the frame of (1), whereas $n$ and $a$ just modify its
absolute value. This fact reinforces the interpretation of $c$, already
given in \cite{Silva} and \cite{Herrera}, in the sense that it represents
the vorticity of the source, when descibed by a rigidly rotating anisotropic
cylinder \cite{Silva} and that it provides the {\it dragging} correction to
the angular velocity of a particle in circular orbits in Lewis spacetime
\cite{Herrera}. However, here there was no need to specify the source that
produces the field and, on the other hand, although some kind of {\it frame
dragging}
effect may also be related to $b$ (see (60) in \cite{Herrera}), this last
parameter does not play any role in the precession under consideration.\par
Another interesting point about (32) is that if $c$ is small, it becomes
\begin{equation}
\Omega\approx\frac{c}{a}r^{(1-n)(n-3)/4},
\end{equation}
and we observe that $n=3$ produces a constant $\Omega$, independent of $r$.
It is known that when $b=c=0$ and $n=3$ the metric becomes locally Taub's
plane metric \cite{Taub,Wang,Silva3}. This fact suggests that the
gravitational potential becomes constant, not modifying the precession with
respect to its distance. \par
Next, introducing
\begin{equation}
\beta\equiv\frac{c}{na},
\end{equation}
we can write (32) as
\begin{equation}
\Omega=\frac{\beta nr^{(1-n)(n-3)/4}}{1-\beta^2r^{2n}}.
\end{equation}
Then performing a series of {\it measurements} of $\Omega$ for different
values of $r$, we can in principle obtain $n$ and $\beta$ by adjusting these
parameters to the obtained curve $\Omega=\Omega(r)$, which in turn allows
for obtaining the value of $c/a$.\par
In the Levi-Civita metric, $b=c=0$, the gyroscope at rest will not precess,
as expected for a vacuum static spacetime (for the electrovac case however,
this may change \cite{Bonnor1}).\par
All these comments above (after equation (32)), are valid as long as we adopt the point of view that eqs.(12)-(13)
globally transform (1) into the Levi-Civita spacetime , if and only if $b=c=0$.
However if we adopt the point of view that only $b$ has to vanish for that
transformation to be
globally valid, then the precession given by (32) is just a coordinate
effect,implying
that there is not frame dragging in the Lewis space-time (Weyl class),
since the precession of the gyroscope
is due to the fact that the original frame of (1) is rotating itself.
This is  quite a surprising result, if we recall
that material sources for (1) consist in steadly rotating fluids (see
\cite{Bon.ref} and \cite{Silva}).Furthermore the vorticity
of the source given in \cite{Silva} is, at the boundary surface,
proportional to $c$ (not $b$).
\section{Precession of gyroscope moving in a circle around the axis of
symmetry}
According to the meaning of $\Omega$ given above, it is clear that the
orientation of the gyroscope, moving around the axis of symmetry, after one
revolution, changes by
\begin{equation}
\Delta\phi^{\prime}=-\Omega\Delta\tau,
\end{equation}
where $\Delta\tau$ is the proper time interval corresponding to one period.
Then from (15),
\begin{equation}
\Delta\phi^{\prime}=-2\pi\frac{\Omega e^{\Psi}}{\omega},
\end{equation}
as measured in the rotating frame. In the original system, we have
\begin{equation}
\Delta\phi=2\pi\left(1-\frac{\Omega e^{\Psi}}{\omega}\right).
\end{equation}
To calculate (24) and (38) for the metric (1) we first obtain from (18) and
(19) using (2)-(5),
\begin{eqnarray}
e^{2\Psi}=aM^2r^{1-n}-\frac{N^2}{n^2a}r^{1+n},\\
\omega_{\phi,r}=2MNe^{-4\Psi}r,
\end{eqnarray}
where
\begin{eqnarray}
M=1+b\omega,\\
N=n\omega-c(1+b\omega).
\end{eqnarray}
Now substituting (39), (40) and (25) into (24) and (38), we obtain,
\begin{eqnarray}
\Omega=MNr^{(1-n^2)/4}\left(M^2ar^{1-n}-\frac{N^2}{n^2a}r^{1+n}\right)^{-1},
\\
\Delta\phi=2\pi\left[1-\frac{MN}{\omega}r^{(1-n^2)/4}\left(M^2ar^{1-n}-\frac
{N^2}{n^2a}r^{1+n}\right)^{-1/2}\right].
\end{eqnarray}
When particle follows a circular geodesic around the axis, then the angular
velocity is (27) and then (43) and (44) become,
\begin{eqnarray}
\Omega=\frac{1}{2}(1-n^2)^{1/2}r^{-(3+n^2)/4},\\
\Delta\phi=2\pi\left\{1-\frac{n(1-n)^{1/2}ar^{-n}}{n(1-n)^{1/2}ar^{-n}
+(1+n)^{1/2}c}\left[\frac{n(1+n)}{2a}\right]^{1/2}r^{-(1-n)^2/4}\right\}.
\end{eqnarray}
Surprisingly neither $b$ nor $c$ enter into the expression (45) for
$\Omega$. Also, the expression (46) is unaffected by $b$. If $\omega_0r\ll
1$ and $c\ll\omega_0$, we have from (46),
\begin{equation}
\Delta\phi\approx\delta+3\pi\frac{\omega_0^2r^2}{a^{5/2}}+
2\pi\frac{c}{\sqrt{a}\omega_0},
\end{equation}
and if $a=1$ we have $\Delta\phi\approx 3\pi\omega_0^2r^2+ 2\pi
c/\omega_0$, which coincides with the Schiff precession \cite{Schiff} in
the Kerr
spacetime, if we identify the Kerr parameter with $c$.
We shall now apply (43)-(46) to some specific cases.
\subsection{Levi-Civita spacetime case}
When in (1) $b=c=0$ we have the static Levi-Civita spacetime then (43) and
(44) reduce to
\begin{eqnarray}
\Omega=\frac{na\omega r^{(1-n)(n-3)/4}}{a^2-\omega^2r^{2n}},\\
\Delta\phi=2\pi\left[1-\frac{n\sqrt{a}r^{-(1-n)^2/4}}
{(a^2-\omega^2r^{2n})^{1/2}}\right].
\end{eqnarray}
Let us now assume that the trajectory of the gyroscope is a geodesic, then
$\omega=\omega_0$ given by
(28), we have from (48) and (49),
\begin{eqnarray}
\Omega=\frac{1}{2}(1-n^2)^{1/2}r^{-(3+n^2)/4},\\
\Delta\phi=2\pi\left\{1-\left[\frac{n(1+n)}{2a}\right]^{1/2}r^{-(1-n)^2/4}
\right\}.
\end{eqnarray}
In the case $\omega_0r\ll 1$, (51) becomes
\begin{equation}
\Delta\phi\approx\delta+3\pi\frac{\omega_0^2r^2}{a^{5/2}},
\end{equation}
and if $a=1$, we have $\Delta\phi\approx 3\pi\omega_0^2r^2$, which coincides
with the Fokker-de Sitter precession \cite{Fokker,Sitter} in the
Schwarzschild spacetime.
It is interesting to note that if $n=0$, which corresponds to the null
circular geodesics \cite{Bonnor,Herrera} as can be seen from (30) when
$W=1$, we have from (49) that $\Delta\phi=2\pi$.
This behaviour means that the precession becomes so large, that
independently of $\omega$, the orientation of the gyroscope is locked to the
lattice of the rotating frame. Exactly the same behaviour appears in the
Schwarzschild spacetime.
\subsection{Flat spacetime case}
From the Cartan scalars we see that only $n$ helps to curve the spacetime
\cite{Silva}. When $n=1$ the spacetime becomes flat and (43) and (44)
become,
\begin{eqnarray}
\Omega=\frac{a\tilde{\omega}}{a^2-\tilde{\omega}^2r^2},\\
\Delta\phi=2\pi\left[1-\frac{(1+bc)\sqrt{a}\tilde{\omega}}
{(\tilde{\omega}+c)(a^2-\tilde{\omega}^2r^2)^{1/2}}\right],
\end{eqnarray}
where $\tilde{\omega}$ is the effective coordinate rate of rotation,
\begin{equation}
\tilde{\omega}=\frac{\omega}{1+b\omega}-c.
\end{equation}
From (54) we see that $b$ and $c$ affect $\tilde{\omega}$ by increasing
(decreasing) it when they are in the same (opposite) sense of rotation with
respect to $\omega$.
Let us first consider the case $b=c=0$, then (48) and (49) become
\begin{eqnarray}
\Omega=\frac{a\omega}{a^2-\omega^2r^2},\\
\Delta\phi=2\pi\left[1-\frac{\sqrt{a}}{(a^2-\omega^2r^2)^{1/2}}\right].
\end{eqnarray}
These expressions, (56) and (57), put in evidence the influence of $a$ on
the Thomas precession of a gyroscope moving around a string with linear
energy density $\lambda$ given by \cite{Wang,Linet}
\begin{equation}
\lambda=\frac{1}{4}\left(1-\frac{1}{\sqrt{a}}\right).
\end{equation}
We recall that $a$ changes the topological structure of the spacetime,
giving rise to an angular deficit $\delta$ equal to \cite{Dowker}
\begin{equation}
\delta=2\pi\left(1-\frac{1}{\sqrt{a}}\right).
\end{equation}
In the case $\omega r\ll 1$, (57) becomes
\begin{equation}
\Delta\phi\approx\delta-\pi\frac{\omega^2r^2}{a^{5/2}},
\end{equation}
and if $a=1$ we have the usual Thomas precession
$\Delta\phi\approx-\pi\omega^2r^2$.
If $b=0$ and $c\ll\omega$ (54) becomes
\begin{equation}
\Delta\phi\approx2\pi\left[1-\frac{\sqrt{a}}{(a^2-\omega^2r^2)^{1/2}}+\frac
{c}{\omega}\frac{a^2+\omega^2r^2}{(a^2-\omega^2r^2)^{3/2}}\right],
\end{equation}
and if $c=0$ and $b\omega\ll 1$ we have from (54)
\begin{equation}
\Delta\phi\approx
2\pi\left[1-\frac{\sqrt{a}}{(a^2-\omega^2r^2)^{1/2}}+b\omega
\frac{\omega^2r^2}{(a^2-\omega^2r^2)^{3/2}}\right].
\end{equation}
We see from (60) and (62) that if the order of magnitude of $c/\omega$ and
$b\omega$ are equal, $O(c/\omega)=O(b\omega)$, then the contribution of $c$
is larger than $b$ to the precession.
The expression (62) exhibits the modifications on the Thomas precession,
associated with the topological defect created by $b$. It is worth noticing
that a quantum scalar particle moving around a spinning cosmic string,
exhibits a phase factor proportional to $b\sqrt{a}$, an evident reminiscence
of the Aharonov-Bohm effect \cite{Jensen}.
\section{Conclusions}
We have been able to establish a set of expressions linking the parameters
of the Lewis metric to quantities obtained from the observation of
gyroscope's precession. In the particular case of a gyroscope at rest in the
original lattice, the relevance of parameter $c$ is clearly illustrated.
Curiously enough, the parameter $b$ does not enter into the expression of
the angular velocity of precession (see discussion above on this point).
 In the case of the gyroscope rotating
around the axis of symmetry, we obtain that in the Levi-Civita case the
precession vanishes at the photon orbit. A similar result is known to happen
for the Schwarzschild \cite{Rindler} and the Ernst \cite{Ragesh} spacetimes.
However in the general case, $b\neq 0$ and $c\neq 0$, the same result is
observed, which is different from previous results found in stationary
spacetimes \cite{Ragesh1,Ragesh2}. This happens probably due to the fact
already mentioned, that the Weyl class of the Lewis metric, is a rather sui
generis class of stationary metrics, since  it is locally static.
For the special cases with $n=1$, we found how different parameters affect
the Thomas precession, providing at the same time a tool for their {\it
measurement}.
We would like to conclude with the following comment. Except for $n$,
neither of the parameters $a,b$ and $c$ of the Lewis metric enter into the
expressions for the physical components of the Riemann tensor \cite{Silva}.
This implies that they cannot be {\it measured} by means of tidal forces
observations. Therefore gyroscope precession experiments (i.e. experiments
leading to the measuring
of the rate of precession of a gyroscope at rest in the frame of a given
metric and /or
the total precession per revolution of a gyroscope circumventing the source
of such metric) provide a good
alternative for observing those aspects of gravitation not directly related
to the curvature. We have in mind not only  topological deffects, as is the
case here,  but  other
 issues appering in the study of gravity and which are not
 directly related with the value of the physical components of the Riemann
tensor (see for example \cite {Bonnor1},
\cite {deFelice},
\cite {deFeliceII},\cite {Semerak} and references therein).In the case of
real experiments which are
now being contemplated as the GP-B in the solar system, it might in
principle (we are completely ignorant about the accuracy
of such experiment) help to determine what is, among all stationary
solutions, the spacetime associated to a rotating source
(which we expect to be the Kerr metric).

\end{document}